\documentclass[a4paper,reqno]{amsart}
\usepackage{amsmath,amssymb,mathrsfs}
\usepackage{txfonts,bm,pifont}
\usepackage{cite}
\usepackage{float}
\usepackage{stackrel}
\usepackage{subcaption}
\usepackage{graphicx,xcolor}
\usepackage{comment}
\usepackage{enumitem}
\usepackage{longtable}
\usepackage{hyperref}

\newtheorem{theorem}{Theorem}[section]

\newtheorem{remark}[theorem]{Remark}
\theoremstyle{definition}

\numberwithin{equation}{section}
\numberwithin{figure}{section}
\numberwithin{table}{section}
\newcommand{\wutilde}[1]{\vrule depth 0pt width 0pt%
{\raise0.8pt\hbox{$\smash{{\mathop{#1} \limits_{\displaystyle\widetilde{}}}}$}}}
\newcommand{\wuhat}[1]{\vrule depth 0pt width 0pt%
{\raise0.6pt\hbox{$\smash{{\mathop{#1} \limits_{\displaystyle\widehat{}}}}$}}}

\newcommand{\al}{\alpha}
\newcommand{\be}{\beta}

\newcommand{\ga}{\gamma}

\newcommand{\ep}{\epsilon}

\newcommand{\PDE}{P$\Delta$E}
\newcommand{\bbZ}{\mathbb{Z}}

\newcommand{\bbC}{\mathbb{C}}
\newcommand{\bbP}{\mathbb{P}}

\newcommand{\qPD}{\text{\rm$q$-P$_{\rm III}^{D_7^{(1)}}$}}
\newcommand{\qP}[1]{\text{\rm$q$-P$_{\rm #1}$}}

\makeatletter
\long\def\@makecaption#1#2{
 \vskip 10pt
 \setbox\@tempboxa\hbox{#1. #2}
 \ifdim \wd\@tempboxa >\hsize #1. #2\par \else \hbox
to\hsize{\hfil\box\@tempboxa\hfil}
 \fi}
\makeatother
\makeatletter
\@namedef{subjclassname@2020}{%
  \textup{2020} Mathematics Subject Classification}
\makeatother
\begin{document}
\allowdisplaybreaks

\title[]{Discrete Painlev\'e transcendent solutions to the multiplicative type discrete KdV equations}
\author{Nobutaka Nakazono}
\address{Institute of Engineering, Tokyo University of Agriculture and Technology, 2-24-16 Nakacho Koganei, Tokyo 184-8588, Japan.}
\email{nakazono@go.tuat.ac.jp}
\begin{abstract}
Hirota's discrete KdV equation is an integrable partial difference equation on $\bbZ^2$, which approaches the Korteweg-de Vries (KdV) equation in a continuum limit. 
In this paper, we show that its multiplicative-discrete versions have the special solutions given by the solutions of $q$-Painlev\'e equations of types $A_J^{(1)}$ $(J=3,4,5,6)$.
\end{abstract}

\maketitle

\section{Introduction}\label{Introduction}
The Korteweg-de Vries (KdV) equation\cite{korteweg1895xli}:
\begin{equation}\label{eqn:kdv}
 u_t+6uu_x+u_{xxx}=0,
\end{equation}
where $u=u(t,x)\in\bbC$ and $(t,x)\in\bbC^2$,
is known as a mathematical model of waves on shallow water surfaces.
The KdV equation is an important equation that has been studied extensively in physics, engineering and mathematics, especially in the field of integrable systems (see, {\it i.e.}, \cite{arbarello2002sketches,drazin1989solitons} and references therein).
In 1977, Hirota found the following integrable discrete version of the KdV (dKdV) equation \cite{HirotaR1977:MR0460934}:
\begin{equation}\label{eqn:dkdv_u_lm}
 u_{l+1,m+1}-u_{l,m}=\dfrac{~1~}{u_{l,m+1}}-\dfrac{~1~}{u_{l+1,m}},
\end{equation}
where $u_{l,m}\in \bbC$ and $(l, m)\in\bbZ^2$.
Indeed, Equation \eqref{eqn:dkdv_u_lm} 
has the similar properties with the KdV equation (e.g., soliton solution, Lax pair and so on) and
approaches Equation \eqref{eqn:kdv} in a continuum limit.
Moreover, in 1991, Capel {\it et al.} found the non-autonomous generation of the  dKdV equation
\cite{CNP1991:MR1111648,TGR2001:REMBLAY2001319,kajiwara2008bilinearization,matsuura2007book}:
\begin{equation}\label{eqn:mul_nonauto_dKdV}
 u_{l+1,m+1}-u_{l,m}
 =\dfrac{q_{m+1}-p_{l}}{u_{l,m+1}}-\dfrac{q_{m}-p_{l+1}}{u_{l+1,m}},
\end{equation}
where $p_l,q_m\in\bbC$ are arbitrary functions of $l$ and $m$, respectively. 

In this paper, we focus on the two multiplicative-discrete type {\PDE}s.
One is
\begin{equation}\label{eqn:mul_dKdV_1}
 u_{l+1,m+1}-u_{l,m}
 =\dfrac{\be_{m+1}-\al_{l}}{u_{l,m+1}}-\dfrac{\be_{m}-\al_{l+1}}{u_{l+1,m}},
\end{equation}
and the other is
\begin{equation}\label{eqn:mul_dKdV_2}
 u_{l+1,m+1}-u_{l,m}=\dfrac{B_{m+1}-A_{l}}{u_{l,m+1}}-\dfrac{B_{m}-A_{l+1}}{u_{l+1,m}},
\end{equation}
where
\begin{equation}
 A_l=\dfrac{(1-\al_l)(\ga-\al_l)}{\al_l},\quad
 B_m=\dfrac{(1-\be_m)(1-\ga\,\be_m)}{\be_m}.
\end{equation}
Here, 
\begin{equation}
 \al_l=\ep^l\al_0,\quad
 \be_m=\ep^m\be_0,
\end{equation}
and $\al_0,\be_0,\ga\in\bbC$ and $\ep\in\bbC^{\ast}$ are parameters.
Each of equations \eqref{eqn:mul_dKdV_1} and \eqref{eqn:mul_dKdV_2} is a special case of Equation \eqref{eqn:mul_nonauto_dKdV}.
Therefore, these are identified as multiplicative type dKdV equations. 

\begin{remark}
As multiplicative-discrete type difference equations are also called $q$-difference equations, it is common to use the parameter $q$ for their shift parameters. 
However, in this paper, the parameter $q$ is used to denote the shift parameters of $q$-Painlev\'e equations, and we also consider the correspondence between the shift parameters of the $q$-Painlev\'e equations and those of the multiplicative dKdV equations.
Therefore, we use the parameter $\ep$ instead of the parameter $q$ for the multiplicative dKdV equations \eqref{eqn:mul_dKdV_1} and \eqref{eqn:mul_dKdV_2} to avoid confusion.
\end{remark}

In this paper, we study the special solutions of Equations \eqref{eqn:mul_dKdV_1} and \eqref{eqn:mul_dKdV_2}.
The distinctive feature of the special solutions given in this paper is that along each of the directions $l\in\bbZ$ and $m\in\bbZ$ these solutions are expressed in terms of discrete Painlev\'e transcendents.
(See Theorems \ref{theorem:A6}--\ref{theorem:A3}.)
The motivation for the discovery of such solutions is as follows:
\begin{description}
\item[\cite{kajiwara2008bilinearization,joshi2021threedimensional}]
the discrete Miura transformation between the non-autonomous dKdV equation \eqref{eqn:mul_nonauto_dKdV} and the lattice modified KdV (lmKdV) equation\cite{NC1995:MR1329559,NQC1983:MR719638,ABS2003:MR1962121}:
\begin{equation}\label{eqn:intro_lmkdv}
 \dfrac{u_{l+1,m+1}}{u_{l,m}}
 =\dfrac{p_l u_{l+1,m}-q_mu_{l,m+1}}{p_lu_{l,m+1}-q_mu_{l+1,m}},
\end{equation}
where $p_l,q_m\in\bbC$ are arbitrary functions of $l$ and $m$, respectively;
\item[\cite{JNS2016:MR3584386,JNS2015:MR3403054,KNY2002:MR1958118,TakenawaT2003:MR1996297}] 
relations between the lmKdV equation \eqref{eqn:intro_lmkdv} and the $\tau$-functions of the discrete Painlev\'e equations.
\end{description}

\subsection{$q$-Painlev\'e equations}
In this subsection, we briefly explain the discrete Painlev\'e equations, especially $q$-Painlev\'e equations.

In the early 20th-century, in order to find new transcendental functions, 
Painlev\'e and Gambier classified all the ordinary differential equations of the type
\begin{equation}
 u_{xx}=F(u,u_x;x),
\end{equation}
where $F$ is a function meromorphic in $x$ and rational in $u$ and $u_x$,
with the Painlev\'e property (the locations of possible branch points and essential singurarities of the solution do not depend on the initial data)\cite{PainleveP1902:MR1554937,GambierB1910:MR1555055}.
As a result, they obtained six new equations.
The resulting equations are now collectively referred to as the {\it Painlev\'e equations}, and the solutions of the Painlev\'e equations are referred to as the {\it Painlev\'e transcendents}.
The Painlev\'e equations are numbered beginning with one: P$_{\rm I}$, $\dots$, P$_{\rm VI}$,
and starting from P$_{\rm VI}$ we can through appropriate limiting processes obtain P$_{\rm J}$ $({\rm J}={\rm I},\dots,{\rm V})$.
Note that P$_{\rm VI}$ was found by Fuchs \cite{FuchsR1905:quelques} before Painlev\'e {\it et al.}

Discrete Painlev\'e equations are nonlinear ordinary difference equations of second order, 
which include discrete analogues of the Painlev\'e equations.
There are only six Painlev\'e equations, but there are an infinite number of discrete Painlev\'e equations.
Moreover, there are three discrete types: elliptic-, multiplicative- and additive- types.
Discrete Painlev\'e equations of the multiplicative-type are especially referred to as {\it $q$-Painlev\'e equations}.
In a similar fashion as the Painlev\'e equations, we also refer to the solutions of discrete Painlev\'e equations as the {\it discrete Painlev\'e transcendents}.

In 2001, Sakai gave the geometric description of discrete Painlev\'e equations, based on types of space of initial values \cite{SakaiH2001:MR1882403,KNY2017:MR3609039}.
The spaces of initial values are constructed by the blow up of $\bbP^2$ at nine base points
({\it i.e.} points where the system is ill defined because it approaches $0/0$).
$q$-Painlev\'e equations are classified into 9 types: 
$A_0^{(1)\ast}$,
$A_1^{(1)}$, \dots, $A_7^{(1)}$,
$A_7^{(1)'}$
according to the configuration of the base points,
and they relate to the (extended) affine Weyl group of the types 
$E_8^{(1)}$,
$E_7^{(1)}$,
$E_6^{(1)}$,
$D_5^{(1)}$,
$A_4^{(1)}$,
$(A_2+A_1)^{(1)}$,
$(A_1+A_1)^{(1)}$,
$A_1^{(1)}$,
$A_1^{(1)}$,
respectively.
These responses for $q$-Painlev\'e equations of types $A_J^{(1)}$ $(J=3,4,5,6)$ are presented specifically in \S \ref{section:A6_proof} and \S \ref{section:A543_proofs}.
Moreover, some typical examples of $q$-Painlev\'e equations of types $A_J^{(1)}$ $(J=3,4,5,6)$ are displayed in Appendix \ref{section:typical_examples_dPs}.

Together with the Painlev\'e equations, the discrete Painlev\'e equations are now regarded as one of the most important classes of equations in the theory of integrable systems (see, e.g., \cite{GR2004:MR2087743,KNY2017:MR3609039}). 
From the point of view of special function, using the Painlev\'e/discrete Painlev\'e transcendents for solving other differential/difference equations is as important as investigating their properties.
\subsection{Main results}\label{subsection:main_result}
In this subsection, we show the main results of this paper.
Theorems \ref{theorem:A6}--\ref{theorem:A4} give special solutions of Equation \eqref{eqn:mul_dKdV_1}, while Theorem \ref{theorem:A3} gives a special solution of Equation \eqref{eqn:mul_dKdV_2}.

\begin{theorem}[$A_6^{(1)}$-type]\label{theorem:A6}
The multiplicative dKdV equation \eqref{eqn:mul_dKdV_1} has the special solution
\begin{equation}
 u_{l,m}=\dfrac{\ep^{3/4}{\be_{m}}^{1/2}(\al_{l}-\be_{m})}{{\al_{l}}^{1/2}(h_{l,m}+\be_{m+2})g_{l,m}}.
\end{equation}
Here, the functions $g_{l,m}$ and $h_{l,m}$ satisfy the system of ordinary difference equations for the $l$-direction:
\begin{equation}\label{eqn:theo_A6_l-direction}
 h_{l+1,m}h_{l,m}=\dfrac{\be_{m+2}(1+g_{l,m})}{{g_{l,m}}^2},\quad
 g_{l+1,m}g_{l,m}=\dfrac{\be_{m}(h_{l+1,m}+\al_{l+3})}{\al_{l+1}h_{l+1,m}(h_{l+1,m}+\be_{m+2})},
\end{equation}
and that for the $m$-direction:
\begin{equation}\label{eqn:theo_A6_m-direction}
 f_{l,m+1}f_{l,m}=\dfrac{\al_l (h_{l,m}+\be_{m+2})}{\be_mh_{l,m}(h_{l,m}+\al_{l+2})},\quad
 h_{l,m+1}h_{l,m}=\dfrac{\al_{l+2}(1+f_{l,m+1})}{{f_{l,m+1}}^2},
\end{equation}
where 
\begin{equation}
 f_{l,m}=\dfrac{1}{g_{l,m}h_{l,m}}.
\end{equation}
Note that each of equations \eqref{eqn:theo_A6_l-direction} and \eqref{eqn:theo_A6_m-direction} is equivalent to the \qPD \eqref{eqn:qp_appendix_A6_2}, which is a $q$-Painlev\'e equation of $A_6^{(1)}$-type. 
(See Appendix \ref{section:typical_examples_dPs} for details.)
\end{theorem}

The proof of Theorem  \ref{theorem:A6} is given in \S \ref{section:A6_proof}.

\begin{theorem}[$A_5^{(1)}$-type]\label{theorem:A5}
The multiplicative dKdV equation \eqref{eqn:mul_dKdV_1} has the special solution
\begin{equation}
 u_{l,m}=\dfrac{d(\al_{l}-\be_{m})}{({\al_{l}}^{1/2}+{\be_{m}}^{1/2}g_{l,m})\,h_{l,m}},
\end{equation}
where $d\in\bbC^\ast$ is an arbitrary parameter.
Here, the functions $g_{l,m}$ and $h_{l,m}$ satisfy 
the system of ordinary difference equations for the $l$-direction:
\begin{equation}\label{eqn:theo_A5_l-direction}
 g_{l+1,m}g_{l,m}
 =\dfrac{d^2({\al_{l}}^{1/2}+f_{l,m})}{f_{l,m}(1+{\al_{l}}^{1/2}f_{l,m})},\quad
 f_{l+1,m}f_{l,m}
 =\dfrac{d^2({\al_{l+1}}^{1/2}+{\be_{m}}^{1/2}g_{l+1,m})}{g_{l+1,m}({\be_{m}}^{1/2}+{\al_{l+1}}^{1/2}g_{l+1,m})},
\end{equation}
where 
\begin{equation}
 f_{l,m}=\dfrac{d^2}{g_{l,m}h_{l,m}},
\end{equation}
and that for the $m$-direction:
\begin{equation}\label{eqn:theo_A5_m-direction}
 h_{l,m+1}h_{l,m}
 =\dfrac{d^2({\be_{m}}^{1/2}+{\al_{l}}^{1/2}g_{l,m})}{g_{l,m}({\al_{l}}^{1/2}+{\be_{m}}^{1/2}g_{l,m})},\quad
 g_{l,m+1}g_{l,m}
 =\dfrac{d^2({\be_{m}}^{1/2}+h_{l,m+1})}{h_{l,m+1}(1+{\be_{m}}^{1/2}h_{l,m+1})}.
\end{equation}
Note that each of equations \eqref{eqn:theo_A5_l-direction} and \eqref{eqn:theo_A5_m-direction} is equivalent to the \qP{III} \eqref{eqn:qp3_appendix_A5}, which is a $q$-Painlev\'e equation of $A_5^{(1)}$-type. 
(See Appendix \ref{section:typical_examples_dPs} for details.)
\end{theorem}

The proof of Theorem  \ref{theorem:A5} is given in \S \ref{subsection:A5_proof}.

\begin{theorem}[$A_4^{(1)}$-type]\label{theorem:A4}
The multiplicative dKdV equation \eqref{eqn:mul_dKdV_1} has the special solution
\begin{equation}
 u_{l,m}=\dfrac{{\be_{m}}^{1/2}(\al_{l}-\be_{m})h_{l,m}}{\ep^{1/4}{\al_{l}}^{3/2}d_1d_2(1+\be_{m} g_{l,m})},
\end{equation}
where $d_1,d_2\in\bbC^\ast$ are arbitrary parameters.
Here, the functions $g_{l,m}$ and $h_{l,m}$ satisfy 
the system of ordinary difference equations for the $l$-direction:
\begin{equation}\label{eqn:theo_A4_l-direction}
\begin{cases}
 g_{l+1,m}g_{l,m}=\dfrac{(\be_{m-1}{d_1}^2{d_2}^2+\al_lf_{l,m})(\be_m+\al_lf_{l,m})}{{\al_l}^2{\be_m}^2{d_2}^2({d_1}^2\be_m+f_{l,m})},\\[1em]
 f_{l+1,m}f_{l,m}=\dfrac{{\be_{m}}^2{d_1}^2{d_2}^2(1+\al_{l+1}g_{l+1,m})(1+\al_{l+1}\,\be_m\, g_{l+1,m})}{{\al_{l+1}}^2(1+\be_m g_{l+1,m})},
\end{cases}
\end{equation}
where
\begin{equation}
 f_{l,m}=\dfrac{{d_1}^2{d_2}^2(1+\al_l\,\be_m\,g_{l,m})}{h_{l,m}},
\end{equation}
and that for the $m$-direction:
\begin{equation}\label{eqn:theo_A4_m-direction}
\begin{cases}
 h_{l,m+1}h_{l,m}=\dfrac{{\al_l}^2{d_1}^2{d_2}^2(1+\be_m\,g_{l,m})(1+\al_l\,\be_m\,g_{l,m})}{{\be_m}^2(1+\al_l\,g_{l,m})},\\[1em]
 g_{l,m+1}g_{l,m}=\dfrac{(\al_{l+1}+\be_{m+1}h_{l,m+1})(\al_l\,{d_1}^2{d_2}^2+\be_{m+1}h_{l,m+1})}{{\al_l}^2\,{\be_{m+1}}^2{d_2}^2(\al_l {d_1}^2+h_{l,m+1})}.
\end{cases}
\end{equation}
Note that each of equations \eqref{eqn:theo_A4_l-direction} and \eqref{eqn:theo_A4_m-direction} is equivalent to the \qP{V} \eqref{eqn:qP_appendix_A4}, which is a $q$-Painlev\'e equation of $A_4^{(1)}$-type. 
(See Appendix \ref{section:typical_examples_dPs} for details.)
\end{theorem}

The proof of Theorem  \ref{theorem:A4} is given in \S \ref{subsection:A4_proof}.

\begin{theorem}[$A_3^{(1)}$-type]\label{theorem:A3}
The multiplicative dKdV equation \eqref{eqn:mul_dKdV_2} has the special solution
\begin{equation}
 u_{l,m}
 =\cfrac{\ep^{1/4}(\be_m\ga-\al_l)\left(f_{l,m}(d_1+x_{l,m})+{d_2}^{1/2}(1+d_1{d_2}^{-1}\,\be_mx_{l,m})\right)}
 {{\al_l}^{1/2}{\be_m}^{1/2}\ga^{1/2}x_{l,m}\left(\be_mf_{l,m}(d_1+x_{l,m})+{d_2}^{1/2}(1+d_1{d_2}^{-1}\,\be_m x_{l,m})\right)},
\end{equation}
where $d_1,d_2\in\bbC^\ast$ are arbitrary parameters.
Here, the functions $f_{l,m}$ and $x_{l,m}$ satisfy 
the system of ordinary difference equations for the $l$-direction:
\begin{equation}\label{eqn:theo_A3_l-direction}
 \begin{cases}
 g_{l+1,m}g_{l,m}=\dfrac{(f_{l,m}+{d_1}^{-1}{d_2}^{1/2}{\al_l}^{-1}\ga)(f_{l,m}+d_1{d_2}^{-1/2}\,{\al_l}^{-1})}{(f_{l,m}+{d_1}^{-1}{d_2}^{1/2})(f_{l,m}+d_1\,{d_2}^{-1/2})},\\[1em]
 f_{l+1,m}f_{l,m}=\dfrac{(g_{l+1,m}+{d_2}^{-1/2}{\al_l}^{-1}{\be_{m-1}}^{1/2}\ga^{1/2})(g_{l+1,m}+{d_2}^{1/2}{\al_l}^{-1}{\be_{m+1}}^{-1/2}\ga^{1/2})}{(g_{l+1,m}+{d_2}^{-1/2}{\be_{m-1}}^{-1/2}\ga^{-1/2})(g_{l+1,m}+{d_2}^{1/2}{\be_{m-1}}^{1/2}\ga^{1/2})},
 \end{cases}
\end{equation}
where
\begin{equation}
 g_{l,m}=\dfrac{{d_2}^{1/2}\ga+d_1\,\al_l\,f_{l,m}}{\al_l\,{\be_{m-1}}^{1/2}\ga^{1/2}(d_1+{d_2}^{1/2}f_{l,m})x_{l,m}},
\end{equation}
and that for the $m$-direction:
\begin{equation}\label{eqn:theo_A3_m-direction}
 \begin{cases}
  y_{l,m+1}y_{l,m}=\dfrac{(x_{l,m}+{d_1}^{-1}{d_2}\,{\be_m}^{-1})(x_{l,m}+d_1{d_2}^{-1}\, {\be_{m-1}}^{-1}\ga^{-1})}{(x_{l,m}+d_1)(x_{l,m}+{d_1}^{-1})},\\[1em]
 x_{l,m+1}x_{l,m}=\dfrac{(y_{l,m+1}+{\al_l}^{-1/2}{\be_m}^{-1})(y_{l,m+1}+{\al_l}^{1/2}{\be_m}^{-1}\ga^{-1})}{(y_{l,m+1}+{\al_l}^{1/2})(y_{l,m+1}+{\al_l}^{-1/2})},
 \end{cases}
\end{equation}
where
\begin{equation}
 y_{l,m}=\dfrac{{\al_l}^{1/2}f_{l,m}(d_1+{d_2}\,\be_{m-1}\,\ga\, x_{l,m})}{{d_2}^{1/2}\be_{m-1}\,\ga(1+d_1\, x_{l,m})}.
\end{equation}
Note that each of equations \eqref{eqn:theo_A3_l-direction} and \eqref{eqn:theo_A3_m-direction} is equivalent to the \qP{VI} \eqref{eqn:qP_appendix_A3}, which is a $q$-Painlev\'e equation of $A_3^{(1)}$-type. 
(See Appendix \ref{section:typical_examples_dPs} for details.)
\end{theorem}

The proof of Theorem  \ref{theorem:A3} is given in \S \ref{subsection:A3_proof}.

\subsection{Plan of the paper}
This paper is organized as follows.
In \S \ref{section:A6_proof}, using the birational representation of the extended affine Weyl group of type $(A_1+A_1)^{(1)}$, we give a proof of Theorem \ref{theorem:A6}.
In \S \ref{section:A543_proofs}, using exactly the same process in \S \ref{section:A6_proof} we prove Theorems \ref{theorem:A5}--\ref{theorem:A3}.
Some concluding remarks are given in \S \ref{ConcludingRemarks}.
In Appendix \ref{section:typical_examples_dPs}, we list some typical examples of $q$-Painlev\'e equations of types $A_J^{(1)}$ ($J=3,4,5,6$) and give the correspondences between those $q$-Painlev\'e equations and the $q$-Painlev\'e equations in Theorems \ref{theorem:A6}--\ref{theorem:A3}.
\section{Proof of Theorem \ref{theorem:A6}}\label{section:A6_proof}
In this section, using the birational representation of the extended affine Weyl group of type $(A_1+A_1)^{(1)}$, which gives rise to the $A_6^{(1)}$-type $q$-Painlev\'e equations \eqref{eqn:qp2_appendix_A6} and \eqref{eqn:qp_appendix_A6_2}, we give a proof of Theorem \ref{theorem:A6}.

\subsection{Birational action of $\widetilde{W}((A_1+A_1')^{(1)})$}\label{subsection:A6_qP_funs}
In this subsection, we show the birational actions of transformation group $\widetilde{W}((A_1+A_1')^{(1)})$.

Let  $a_0$, $a_1$, $b$, $q$ be complex parameters  
and $f_0$, $f_1$, $f_2$ be complex variables satisfying
\begin{equation}\label{eqn:conds_af_A1A1}
 a_0a_1=q,\quad
 f_0f_1f_2=1.
\end{equation}
We define the transformation group $\widetilde{W}((A_1+A_1')^{(1)})=\langle s_0,s_1,w_0,w_1,\pi\rangle$ as follows: each element of $\widetilde{W}((A_1+A_1')^{(1)})$ is an isomorphism from the field of rational functions $K(f_0,f_1)$, where $K=\bbC(a_0,a_1,b)$, to itself.
Note that  for each element $w\in\widetilde{W}((A_1+A_1')^{(1)})$ and function $F=F(a_i,b,f_j)$, 
we use the notation $w.F$ to mean $w.F=F(w.a_i,w.b,w.f_j)$, that is, 
$w$ acts on the arguments from the left. 
The actions of $\widetilde{W}((A_1+A_1')^{(1)})$ on the parameters are given by
\begin{align*}
 &s_0:(a_0,a_1,b)
 \mapsto
 \left(\dfrac{1}{a_0}, {a_0}^2 a_1, \dfrac{b}{a_0}\right),
 &&s_1:(a_0,a_1,b)
 \mapsto
 \left(a_0 {a_1}^2, \dfrac{1}{a_1}, a_1 b\right),\\
 &w_0:(a_0,a_1,b,q)
 \mapsto
 \left(\dfrac{1}{a_0}, \dfrac{1}{a_1}, \dfrac{b}{a_0},\frac{1}{q}\right),
 &&w_1:(a_0,a_1,b,q)
 \mapsto
 \left(\dfrac{1}{a_0}, \dfrac{1}{a_1}, \dfrac{b}{{a_0}^2 a_1},\frac{1}{q}\right),\\
 &\pi:(a_0,a_1,b,q)
 \mapsto
 \left(\dfrac{1}{a_1}, \dfrac{1}{a_0}, \dfrac{b}{a_0 a_1},\frac{1}{q}\right),
\end{align*}
while those on the variables are given by
\begin{align*}
 &s_0:(f_0,f_1,f_2)\mapsto
 \left(\dfrac{f_0 (a_0 f_0+a_0+f_1)}{f_0+f_1+1},\dfrac{f_1 (a_0 f_0+f_1+1)}{a_0 (f_0+f_1+1)},
 \dfrac{a_0 f_2(f_0+f_1+1)^2}{(a_0 f_0+a_0+f_1)(a_0 f_0+f_1+1)}\right),\\
 &s_1:(f_0,f_1)\mapsto
 \left(\dfrac{f_0 (a_0 a_1+b f_0 f_1)}{a_1 (a_0+b f_0 f_1)},\dfrac{a_1 f_1 (a_0+b f_0 f_1)}{a_0 a_1+b f_0 f_1}\right),\\
 &w_0:(f_0,f_1,f_2)\mapsto
 \left(\dfrac{a_0 (f_0+1)}{b f_0 f_1},\dfrac{a_0 f_0+a_0+b f_0 f_1}{a_0 b f_0 (f_0+1)},
 \dfrac{b^2 f_0}{f_2 (a_0 f_0+a_0+b f_0 f_1)}\right),\\
 &w_1:(f_0,f_1)\mapsto\left(f_1,f_0\right),\\
 &\pi:(f_1,f_2)\mapsto\left(\dfrac{a_0 (f_0+1)}{b f_0 f_1},\dfrac{b f_1}{a_0(f_0+1)}\right).
\end{align*}

\begin{remark}\label{remark:action_identity}
We follow the convention that the variables and parameters not explicitly including in the actions listed in the equations above are the ones that remain unchanged under the action of the corresponding transformation.
That is, the transformation acts as an identity on those variables and parameters.

\end{remark}

The transformation group $\widetilde{W}((A_1+A_1')^{(1)})$ 
forms the extended affine Weyl group of type $(A_1+A_1)^{(1)}$\cite{SakaiH2001:MR1882403,JNS2015:MR3403054}. 
Namely, the transformations satisfy the following fundamental relations:
\begin{subequations}
\begin{align}
 &{s_0}^2={s_1}^2=(s_0s_1)^\infty=1,\qquad
 {w_0}^2={w_1}^2=(w_0w_1)^\infty=1,\\
 &\pi^2=1,\quad
 \pi s_0=s_1\pi,\quad 
 \pi w_0=w_1\pi,
\end{align}
\end{subequations}
and the action of $W(A_1^{(1)})=\langle s_0,s_1\rangle$ and that of $W(A_1'^{(1)})=\langle w_0,w_1\rangle$ commute.
We note that the relation $(ww')^\infty=1$ for transformations $w$ and $w'$ means that
there is no positive integer $N$ such that $(ww')^N=1$.

We define the translations in $\widetilde{W}((A_1+A_1')^{(1)})$ by
\begin{equation}
 T_1=w_0 w_1,\quad
 T_2=\pi s_1 w_0,\quad
 T_3=\pi s_0 w_0,
\end{equation}
whose actions on the parameters are given by the translational motions:
\begin{subequations}
\begin{align}
 &T_1:(a_0,a_1,b)\mapsto(a_0,a_1,qb),\\
 &T_2:(a_0,a_1,b)\mapsto(qa_0,q^{-1}a_1,b),\\
 &T_3:(a_0,a_1,b)\mapsto(q^{-1}a_0,qa_1,q^{-1}b).
\end{align}
\end{subequations}
Note that the following hold:
\begin{equation}\label{eqn:conds_T_A1A1}
 T_1T_2T_3=1,\quad
 T_iT_j=T_jT_i\quad (i,j=1,2,3),
\end{equation}
and the parameter $q$ is invariant under the action of each translation.

\subsection{Proof of Theorem \ref{theorem:A6}}\label{subsection:A6_dKdV}
In this subsection, using the birational action of $\widetilde{W}((A_1+A_1')^{(1)})$ we give a proof of Theorem \ref{theorem:A6}.

Let us firstly derive the $A_6^{(1)}$-type $q$-Painlev\'e equations \eqref{eqn:qp2_appendix_A6} and \eqref{eqn:qp_appendix_A6_2} from the birational action of $\widetilde{W}((A_1+A_1')^{(1)})$.
Define the $f$-functions by
\begin{equation}
 f_0^{l_1,l_2,l_3}={T_1}^{l_1}{T_2}^{l_2}{T_3}^{l_3}(f_0),\quad
 f_1^{l_1,l_2,l_3}={T_1}^{l_1}{T_2}^{l_2}{T_3}^{l_3}(f_1),\quad
 f_2^{l_1,l_2,l_3}={T_1}^{l_1}{T_2}^{l_2}{T_3}^{l_3}(f_2),
\end{equation}
where $l_1,l_2,l_3\in\bbZ$. 
From the condition in \eqref{eqn:conds_af_A1A1} and relations \eqref{eqn:conds_T_A1A1}, we have
\begin{equation}\label{eqn:cond_f_fun_A6}
 f_0^{l_1,l_2,l_3}f_1^{l_1,l_2,l_3}f_2^{l_1,l_2,l_3}=1,\qquad
 f_i^{l_1,l_2,l_3}=f_i^{0,l_2-l_1,l_3-l_1}\quad (i=1,2,3).
\end{equation}
The $f$-functions satisfy the second order ordinary difference equation for $l_1$-direction:
\begin{equation}\label{eqn:A6_qP_T1}
 \left(f_0^{l_1+1,l_2,l_3}f_0^{l_1,l_2,l_3}-\dfrac{1}{q^{l_1-l_3}b}\right)
 \left(f_0^{l_1-1,l_2,l_3}f_0^{l_1,l_2,l_3}-\dfrac{1}{q^{l_1-l_3-1}b}\right)
 =\dfrac{q^{-l_1-l_2+2l_3}a_1f_0^{l_1,l_2,l_3}}{b(1+f_0^{l_1,l_2,l_3})},
\end{equation}
the system of first order ordinary difference equations for $l_2$-direction:
\begin{equation}\label{eqn:A6_qP_T2}
 \begin{cases}
 f_2^{l_1,l_2+1,l_3}f_2^{l_1,l_2,l_3}
 =\dfrac{q^{l_1-l_3-1}b}{f_1^{l_1,l_2,l_3}(1+f_1^{l_1,l_2,l_3})},\\[1em]
 f_1^{l_1,l_2+1,l_3}f_1^{l_1,l_2,l_3}
 =\dfrac{q^{l_2-l_3}a_0(qf_2^{l_1,l_2+1,l_3}+q^{l_1-l_3}b)}{f_2^{l_1,l_2+1,l_3}(q^{l_2-l_3+1}a_0f_2^{l_1,l_2+1,l_3}+q^{l_1-l_3}b)},
 \end{cases}
\end{equation}
and that for $l_3$-direction:
\begin{equation}\label{eqn:A6_qP_T3}
 \begin{cases}
 f_0^{l_1,l_2,l_3+1}f_0^{l_1,l_2,l_3}
 =\dfrac{qf_2^{l_1,l_2,l_3}+q^{l_1-l_2}a_1b}{f_2^{l_1,l_2,l_3}(qf_2^{l_1,l_2,l_3}+q^{l_1-l_3}b)},\\[1em]
 f_2^{l_1,l_2,l_3+1}f_2^{l_1,l_2,l_3}
 =\dfrac{q^{l_1-l_2-1}a_1b}{f_0^{l_1,l_2,l_3+1}(f_0^{l_1,l_2,l_3+1}+1)}.
 \end{cases}
\end{equation}
Note that Equations \eqref{eqn:A6_qP_T1}, \eqref{eqn:A6_qP_T2} and \eqref{eqn:A6_qP_T3} respectively follow from
\begin{align}
 &T_1(f_0)f_0=\dfrac{1}{b}+\dfrac{f_0f_1}{a_0(f_0+1)},\quad
 {T_1}^{-1}(f_0)f_0=\dfrac{1}{q^{-1}b}+\dfrac{1}{q^{-1}b f_1},\\
 &T_2(f_2)f_2=\dfrac{q^{-1}b}{f_1(1+f_1)},\quad
 T_2(f_1)f_1=\dfrac{a_0 \big(b+q T_2(f_2)\big)}{T_2(f_2)\big(q a_0 T_2(f_2)+b\big)},\\
 &T_3(f_0)f_0=\dfrac{a_1 b+q f_2}{f_2(b+q f_2)},\quad
 T_3(f_2)f_2=\dfrac{q^{-1}a_1 b}{T_3(f_0)\big(T_3(f_0)+1\big)}.
\end{align}

\begin{remark}
Equation \eqref{eqn:A6_qP_T1} is equivalent to the \qP{II} \eqref{eqn:qp2_appendix_A6},
and each of equations \eqref{eqn:A6_qP_T2} and \eqref{eqn:A6_qP_T3} is equivalent to the \qPD \eqref{eqn:qp_appendix_A6_2}.
Indeed, the correspondence between Equations \eqref{eqn:A6_qP_T1} and \eqref{eqn:qp2_appendix_A6} is given by
\begin{equation}
 F=f_0^{l_1,l_2,l_3},\quad
 \overline{F}=f_0^{l_1+1,l_2,l_3},\quad
 \underline{F}=f_0^{l_1-1,l_2,l_3},\quad
 t=q^{l_1-l_3}b,\quad
 c_1=q^{-l_2+l_3}a_1,
\end{equation}
that between Equations \eqref{eqn:A6_qP_T2} and \eqref{eqn:qp_appendix_A6_2} is given by
\begin{equation}
\begin{split}
 &F=\dfrac{1}{f_2^{l_1,l_2,l_3}},\quad
 G=\dfrac{1}{f_1^{l_1,l_2-1,l_3}},\quad
 \overline{F}=\dfrac{1}{f_2^{l_1,l_2+1,l_3}},\quad
 \overline{G}=\dfrac{1}{f_1^{l_1,l_2,l_3}},\\
 &t=q^{-l_1+l_2}a_0b^{-1},\quad
 c_1=q^{-l_1+l_3+1}b^{-1},
\end{split}
\end{equation}
and that between Equations \eqref{eqn:A6_qP_T3} and \eqref{eqn:qp_appendix_A6_2} is given by
\begin{equation}
\begin{split}
 &F=\dfrac{1}{f_2^{l_1,l_2,l_3}},\quad
 G=\dfrac{1}{f_0^{l_1,l_2,l_3}},\quad
 \overline{F}=\dfrac{1}{f_2^{l_1,l_2,l_3+1}},\quad
 \overline{G}=\dfrac{1}{f_0^{l_1,l_2,l_3+1}},\\
 &t=q^{-l_1+l_3+1}b^{-1},\quad
 c_1=q^{-l_1+l_2+1}{a_1}^{-1}b^{-1}.
\end{split}
\end{equation}
\end{remark}

We are now in a position to prove Theorem \ref{theorem:A6}.
Letting
\begin{equation}
 u=\dfrac{1-a_1}{q^{1/4}{a_1}^{1/2}f_0(b+qf_2)},
\end{equation}
we can verify that the following relation holds:
\begin{equation}
 T_2T_3(u)-u=\dfrac{b^{-1}(1-q^{-1}{a_1}^{-1})}{T_3(u)}-\dfrac{b^{-1}(q^{-1}-{a_1}^{-1})}{T_2(u)}.
\end{equation}
Applying ${T_2}^{l}{T_3}^{m}$ to the equation above and setting
\begin{equation}\label{eqn:A6_U_qPA6}
 U_{l,m}
 ={T_2}^{l}{T_3}^{m}(u)
 =\dfrac{1-q^{-l+m}a_1}{q^{(-2l+2m+1)/4}{a_1}^{1/2}f_0^{0,l,m}(q^{-m}b+qf_2^{0,l,m})},
\end{equation}
we obtain
\begin{equation}\label{eqn:A6_U_dKdV}
 U_{l+1,m+1}-U_{l,m}
 =\dfrac{q^mb^{-1}-q^{l-1}{a_1}^{-1}b^{-1}}{U_{l,m+1}}-\dfrac{q^{m-1}b^{-1}-q^{l}{a_1}^{-1}b^{-1}}{U_{l+1,m}}.
\end{equation}
Equation \eqref{eqn:A6_U_dKdV} is equivalent to the multiplicative dKdV equation \eqref{eqn:mul_dKdV_1} and the correspondence is given by the following:
\begin{equation}\label{eqn:corre_dKdV_A6fun}
 u_{l,m}=U_{l,m},\quad
 \al_{l}=q^{l-1}{a_1}^{-1}b^{-1},\quad
 \be_{m}=q^{m-1}b^{-1},\quad
 \ep=q.
\end{equation}

From the way $f_0^{0,l,m}$ and $f_2^{0,l,m}$ are constructed, we know that they are rational functions over $K=\bbC(a_0,a_1,b)$ of the two unknown variables $f_0$ and $f_1$,
that is,
\begin{equation}
 f_0^{0,l,m},f_2^{0,l,m}\in K(f_0,f_1).
\end{equation}
Since the functions $f_0^{0,l,m}$ and $f_2^{0,l,m}$ collectively satisfy Equations \eqref{eqn:A6_qP_T2} and \eqref{eqn:A6_qP_T3} and the number of initial values of a second-order ordinary difference equations is 2, the functions $f_0^{0,l,m}$ and $f_2^{0,l,m}$ give the general solution to each of equations \eqref{eqn:A6_qP_T2} and \eqref{eqn:A6_qP_T3} with $l=l_2-l_1$ and $m=l_3-l_1$. 
Moreover, from the relation \eqref{eqn:A6_U_qPA6}, we find that the functions $f_0^{0,l,m}$ and $f_2^{0,l,m}$ also give the special solutions to Equation \eqref{eqn:A6_U_dKdV}.
Therefore, setting
\begin{equation}
 f_{l,m}=\dfrac{1}{f_0^{0,l,m}},\quad
 g_{l,m}=\dfrac{1}{f_1^{0,l,m}},\quad
 h_{l,m}=\dfrac{1}{f_2^{0,l,m}},
\end{equation}
and using Equations \eqref{eqn:cond_f_fun_A6}, \eqref{eqn:A6_qP_T2}, \eqref{eqn:A6_qP_T3} and \eqref{eqn:A6_U_qPA6} with the correspondence \eqref{eqn:corre_dKdV_A6fun},
we have completed the proof of Theorem \ref{theorem:A6}.

\section{Proofs of Theorems \ref{theorem:A5}--\ref{theorem:A3}}\label{section:A543_proofs}
In this section, using the transformation groups $\widetilde{W}((A_2+A_1)^{(1)})$, $\widetilde{W}(A_4^{(1)})$ and $\widetilde{W}(D_5^{(1)})$,
which respectively relate to $A_5^{(1)}$-, $A_4^{(1)}$- and $A_3^{(1)}$-type $q$-Painlev\'e equations, 
we give proofs of Theorems \ref{theorem:A5}--\ref{theorem:A3}.
Since each process for demonstrating the results is exactly the same as the process in \S \ref{section:A6_proof}, we omit a detailed discussion for brevity.

\subsection{Proof of Theorem \ref{theorem:A5}}\label{subsection:A5_proof}
In this subsection using the transformation group $\widetilde{W}((A_2+A_1)^{(1)})$ we give a proof of Theorem \ref{theorem:A5}.

Let $a_0$, $a_1$, $a_2$, $c$, $q$ be complex parameters and $f_0$, $f_1$, $f_2$ be complex variables satisfying
\begin{equation}\label{eqn:conds_af_A2A1}
 a_0a_1a_2=q,\quad
 f_0f_1f_2=qc^2.
\end{equation}
The transformation group $\widetilde{W}((A_2+A_1)^{(1)})=\langle s_0,s_1,s_2,\pi, w_0,w_1,r\rangle$ is defined by the actions on the parameters:
\begin{equation*}
 s_i(a_j)=a_j{a_i}^{-C_{ij}},\quad 
 \pi(a_i)=a_{i+1},\quad
 w_0(c)=c^{-1},\quad
 w_1(c)=q^{-2}c^{-1},\quad
 r(c)=q^{-1}c^{-1},
\end{equation*}
where $i,j\in\bbZ/3\bbZ$ and $(C_{ij})_{i,j=0}^2$ is the Cartan matrix of type $A_2^{(1)}$:
\begin{equation*}
 (C_{ij})_{i,j=0}^2
 =\left(\begin{array}{ccc}2&-1&-1\\-1&2&-1\\-1&-1&2\end{array}\right)
\end{equation*}
and those on the variables:
\begin{align*}
 &s_i(f_{i-1})=f_{i-1}\dfrac{1+a_if_i}{a_i+f_i},\quad
 s_i(f_i)=f_i,\quad
 s_i(f_{i+1})=f_{i+1}\dfrac{a_i+f_i}{1+a_if_i},\quad
 \pi(f_i) = f_{i+1},\\
 &w_0(f_i)=\dfrac{a_ia_{i+1}(a_{i-1}a_i+a_{i-1}f_i+f_{i-1}f_i)}
  {f_{i-1}(a_ia_{i+1}+a_if_{i+1}+f_if_{i+1})},\\
 &w_1(f_i)=\dfrac{1+a_if_i+a_ia_{i+1}f_if_{i+1}}
  {a_ia_{i+1}f_{i+1}(1+a_{i-1}f_{i-1}+a_{i-1}a_if_{i-1}f_i)},\quad
  r(f_i)={f_i}^{-1},
\end{align*}
where $i\in\bbZ/3\bbZ$.
See Remark \ref{remark:action_identity} for the convention on how to write these actions.
The transformation group $\widetilde{W}((A_2+A_1)^{(1)})$ satisfies the fundamental relations of the extended affine Weyl group of type $(A_2+A_1)^{(1)}$\cite{KNY2001:MR1876614}:
\begin{subequations}
\begin{align}
 &{s_i}^2=(s_is_{i+1})^3=1,
 &&\pi^3=1,\quad
 \pi s_i = s_{i+1}\pi,\\
 &{w_0}^2={w_1}^2=(w_0w_1)^\infty=1,
 &&r^2=1,\quad
 rw_0=w_1r,
\end{align}
\end{subequations}
where $i\in\bbZ/3\bbZ$, and the action of $\widetilde{W}(A_2^{(1)})=\langle s_0,s_1,s_2,\pi\rangle$ and 
that of $\widetilde{W}(A_1^{(1)})=\langle w_0,w_1,r\rangle$ commute.

Define the translations $T_1$ and $T_2$ by
\begin{equation}\label{eqn:def_translations}
 T_1=\pi s_2s_1,\quad
 T_2=\pi s_0s_2,
\end{equation}
whose actions on the parameters are given by
\begin{subequations}
\begin{align}
 &T_1:(a_0,a_1,a_2,c)\mapsto(qa_0,q^{-1}a_1,a_2,c),\\
 &T_2:(a_0,a_1,a_2,c)\mapsto(a_0,qa_1,q^{-1}a_2,c).
\end{align}
\end{subequations}
Note that the translations $T_1$ and $T_2$ commute with each other and the parameter $q$ is invariant under the action of each translation.

Define the $f$-functions by
\begin{equation}\label{eqn:def_f0f1f2_A5}
 f_{0}^{l,m}= {T_1}^l{T_2}^m(f_0),\quad
 f_{1}^{l,m}= {T_1}^l{T_2}^m(f_1),\quad
 f_{2}^{l,m}= {T_1}^l{T_2}^m(f_2),
\end{equation}
where $l,m\in\bbZ$, which from \eqref{eqn:conds_af_A2A1} satisfy
\begin{equation}\label{eqn:cond_f_fun_A5}
 f_{0}^{l,m} f_{1}^{l,m} f_{2}^{l,m}=qc^2.
\end{equation}
The $f$-functions satisfy the following $q$-difference equation for $l$-direction:
\begin{equation}\label{eqn:qP3_lattice}
 f_{1}^{l+1,m} f_{1}^{l,m}
 =\dfrac{qc^2(1+q^la_0f_{0}^{l,m})}{f_{0}^{l,m}(q^la_0+f_{0}^{l,m})},\quad
 f_{0}^{l+1,m} f_{0}^{l,m}
 =\dfrac{qc^2(1+q^{l-m}a_0a_2f_{1}^{l+1,m})}{f_{1}^{l+1,m}(q^{l-m}a_0a_2+f_{1}^{l+1,m})},
\end{equation}
and that for $m$-direction:
\begin{equation}\label{eqn:qP3_latticeT2}
 f_{2}^{l,m+1} f_{2}^{l,m}
 =\dfrac{qc^2(1+q^{m-l}a_1f_{1}^{l,m})}{f_{1}^{l,m}(q^{m-l}a_1+f_{1}^{l,m})},\quad
 f_{1}^{l,m+1} f_{1}^{l,m}
 =\dfrac{qc^2(1+q^ma_0a_1f_{2}^{l,m+1})}{f_{2}^{l,m+1}(q^ma_0a_1+f_{2}^{l,m+1})}.
\end{equation}
Note that Equations \eqref{eqn:qP3_lattice} and \eqref{eqn:qP3_latticeT2} follow from
\begin{align}
 & T_1(f_1)=\dfrac{qc^2(1+a_0f_0)}{f_1f_0(a_0+f_0)},\quad
 {T_1}^{-1}(f_0)=\dfrac{qc^2(1+q^{-1}a_0a_2f_1)}{f_0f_1(q^{-1}a_0a_2+f_1)},\\
 &T_2(f_2)=\dfrac{qc^2(1+a_1f_1)}{f_2f_1(a_1+f_1)},\quad
 {T_2}^{-1}(f_1)=\dfrac{qc^2(1+q^{-1}a_0a_1f_2)}{f_1f_2(q^{-1}a_0a_1+f_2)},
\end{align}
respectively.
\begin{remark}
Each of equations \eqref{eqn:qP3_lattice} and \eqref{eqn:qP3_latticeT2} is equivalent to the \qP{III} \eqref{eqn:qp3_appendix_A5}.
The correspondence between Equations \eqref{eqn:qP3_lattice} and \eqref{eqn:qp3_appendix_A5} is given by
\begin{equation}
\begin{split}
 &F=f_{0}^{l,m},\quad
 G=f_{1}^{l,m},\quad
 \overline{F}=f_{0}^{l+1,m},\quad
 \overline{G}=f_{1}^{l+1,m},\\
 &t=q^la_0,\quad
 c_1=qc^2,\quad
 c_2=q^{-m}a_2,
\end{split}
\end{equation}
while that between Equations \eqref{eqn:qP3_latticeT2} and \eqref{eqn:qp3_appendix_A5} is given by
\begin{equation}
\begin{split}
 &F=f_{1}^{l,m},\quad
 G=f_{2}^{l,m},\quad
 \overline{F}=f_{1}^{l,m+1},\quad
 \overline{G}=f_{2}^{l,m+1},\\
 &t=q^{m-l}a_1,\quad
 c_1=qc^2,\quad
 c_2=q^la_0.
\end{split}
\end{equation}
\end{remark}

Letting
\begin{equation}\label{eqn:A5_u_qPA5}
 u=\dfrac{q^{1/2}c({a_0}^{-2}-q^{-2}{a_2}^2)}{({a_0}^{-1}+q^{-1}a_2f_1)f_2},
\end{equation}
we can verify that the following relation holds:
\begin{equation}
 T_1T_2(u)-u=\dfrac{q^{-4}{a_2}^2-{a_0}^{-2}}{T_2(u)}-\dfrac{q^{-2}{a_2}^2-q^{-2}{a_0}^{-2}}{T_1(u)}.
\end{equation}
Therefore, applying ${T_1}^{l}{T_2}^{m}$ to the equation above
we obtain 
\begin{equation}
 U_{l+1,m+1}-U_{l,m}
 =\dfrac{q^{-2m-4}{a_2}^2-q^{-2l}{a_0}^{-2}}{U_{l,m+1}}-\dfrac{q^{-2m-2}{a_2}^2-q^{-2l-2}{a_0}^{-2}}{U_{l+1,m}},
\end{equation}
where $U_{l,m}={T_1}^{l}{T_2}^{m}(u)$, which is equivalent to the multiplicative dKdV equation \eqref{eqn:mul_dKdV_1} with the following correspondence:
\begin{equation}\label{eqn:corre_dKdV_A5fun}
 u_{l,m}=U_{l,m},\quad
 \al_{l}=q^{-2l}{a_0}^{-2},\quad
 \be_{m}=q^{-2m-2}{a_2}^2,\quad
 \ep=q^{-2}.
\end{equation}
Therefore, setting
\begin{equation}
 d=\ep^{-1/4}c,\quad
 f_{l,m}=f_{0}^{l,m},\quad
 g_{l,m}=f_{1}^{l,m},\quad
 h_{l,m}=f_{2}^{l,m},
\end{equation}
and using Equations \eqref{eqn:cond_f_fun_A5}, \eqref{eqn:qP3_lattice}, \eqref{eqn:qP3_latticeT2} and \eqref{eqn:A5_u_qPA5} with the correspondence \eqref{eqn:corre_dKdV_A5fun},
we have completed the proof of Theorem \ref{theorem:A5}.

\subsection{Proof of Theorem \ref{theorem:A4}}\label{subsection:A4_proof}
In this subsection using the transformation group $\widetilde{W}(A_4^{(1)})$ we give a proof of Theorem \ref{theorem:A4}.

Let $a_i$ $(i=0,\dots,4)$ and $q$ be complex parameters satisfying
\begin{equation}
 a_0a_1a_2a_3a_4=q,
\end{equation}
and $f_i^{(j)}$ $(i=1,2,~j=1,\dots,5)$ be complex variables satisfying
\begin{subequations}\label{eqns:conds_f_vari_A4}
\begin{align}
 &f_2^{(j)}=\dfrac{a_ja_{j+1}(a_{j+2}a_{j+3}+a_jf_1^{(j+3)})}{{a_{j+3}}^2f_1^{(j+1)}},
 &&a_4{a_0}^2f_1^{(2)}f_1^{(3)}=a_2a_3(a_0+a_2f_1^{(5)}),\\
 &a_0{a_1}^2f_1^{(3)}f_1^{(4)}=a_3a_4(a_1+a_3f_1^{(1)}),
 &&a_2{a_3}^2f_1^{(5)}f_1^{(1)}=a_0a_1(a_3+a_0f_1^{(3)}),\label{eqn:conds_f_vari_A4_2}
\end{align}
\end{subequations}
where  $j\in\bbZ/5\bbZ$.
Note that the number of $f$-variables is essentially two.
Indeed, using the relations \eqref{eqns:conds_f_vari_A4} we can express all $f$-variables only by $f_1^{(1)}$ and $f_1^{(3)}$.
The transformation group $\widetilde{W}(A_4^{(1)})=\langle s_0,s_1,s_2,s_3,s_4,\sigma,\iota\rangle$ is defined by the actions on the parameters:
\begin{align*}
 &s_i(a_j)=a_j{a_i}^{-C_{ij}},\quad 
 \sigma(a_i)=a_{i+1},\\
 &\iota:(a_0,a_1,a_2,a_3,a_4,q)\mapsto ({a_0}^{-1},{a_4}^{-1},{a_3}^{-1},{a_2}^{-1},{a_1}^{-1},q^{-1}),
\end{align*}
where $i,j\in\bbZ/5\bbZ$ and $(C_{ij})_{i,j=0}^4$ is the Cartan matrix of type $A_4^{(1)}$:
\begin{equation*}
 (C_{ij})_{i,j=0}^4
 =\left(\begin{array}{ccccc}
 2&-1&0&0&-1\\-1&2&-1&0&0\\0&-1&2&-1&0\\0&0&-1&2&-1\\-1&0&0&-1&2\end{array}\right)
\end{equation*}
and those on the variables: 
\begin{align*}
 &s_j(f_1^{(j+3)})=f_2^{(j+3)},\quad
 s_j(f_2^{(j+3)})=f_1^{(j+3)},\quad
 s_j(f_1^{(j)})=\dfrac{a_{j+4}(a_{j+2}+a_ja_{j+4}f_1^{(j+2)})}{(a_ja_{j+1}{a_{j+2}}^2)f_1^{(j+4)}},\\
 &s_j(f_2^{(j+2)})=\dfrac{a_ja_{j+3}a_{j+4}(a_{j+2}+a_ja_{j+4}f_1^{(j+2)}+a_ja_{j+1}a_{j+2}f_1^{(j+4)})}{a_{j+1}f_1^{(j+4)}f_2^{(j+3)}},\\
 &s_j(f_2^{(j+4)})=\dfrac{a_ja_{j+1}{a_{j+2}}^2f_1^{(j+4)}f_1^{(j)}f_2^{(j+4)}}{a_{j+4}(a_{j+2}+a_ja_{j+4}f_1^{(j+2)})},\\
 &s_j(f_2^{(j)})=\dfrac{a_ja_{j+1}a_{j+4}+a_{j+3}a_{j+4}f_1^{(j+1)}+a_j{a_{j+1}}^2a_{j+2}f_1^{(j+4)}}{a_ja_{j+1}a_{j+3}f_1^{(j+1)}f_1^{(j+4)}},\quad
 \pi(f_1^{(j)})=f_1^{(j+1)},\\
 &\pi(f_2^{(j)})=f_2^{(j+1)},\quad
 \iota(f_1^{(j)})=f_1^{(3-j)},\quad
 \iota(f_2^{(j)})=\dfrac{a_{2-j}(a_{5-j}+a_{2-j}a_{3-j}f_1^{(5-j)})}{a_{3-j}a_{4-j}{a_{5-j}}^2f_1^{(2-j)}},
\end{align*}
where $j\in\bbZ/5\bbZ$.
See Remark \ref{remark:action_identity} for the convention on how to write these actions.
The transformation group $\widetilde{W}(A_4^{(1)})$ satisfies the fundamental relations of the extended affine Weyl group of type $A_4^{(1)}$\cite{JNS2016:MR3584386,TsudaT2006:MR2207047}:
\begin{subequations}
\begin{align}
 &{s_i}^2=1,\quad (s_is_{i\pm 1})^3=1,\quad
 (s_is_j)^2=1,\quad j\neq i\pm 1,\\
 &\sigma^5=1,\quad
 \sigma s_i=s_{i+1}\sigma,\quad
 \iota^2=1,\quad 
 \iota s_{\{0,1,2,3,4\}}=s_{\{0,4,3,2,1\}}\iota,
\end{align}
\end{subequations}
where $i,j\in\bbZ/5\bbZ$.

Define the translations $T_1$ and $T_2$ by
\begin{equation}
 T_1=\sigma s_4s_3s_2s_1,\quad
 T_2=\sigma s_1s_0s_4s_3,
\end{equation}
whose actions on the parameters are given by
\begin{subequations}
\begin{align}
 &T_1:(a_0,a_1,a_2,a_3,a_4)\mapsto(qa_0,q^{-1}a_1,a_2,a_3,a_4),\\
 &T_2:(a_0,a_1,a_2,a_3,a_4)\mapsto(a_0,a_1,qa_2,q^{-1}a_3,a_4).
\end{align}
\end{subequations}
Note that the translations $T_1$ and $T_2$ commute with each other and the parameter $q$ is invariant under the action of each translation.

Define the $f$-functions by
\begin{equation}
 f_{l,m}^{(1)}={T_1}^l{T_2}^m(f_1^{(1)}),\quad
 f_{l,m}^{(3)}={T_1}^l{T_2}^m(f_1^{(3)}),\quad
 f_{l,m}^{(5)}={T_1}^l{T_2}^m(f_1^{(5)}),
\end{equation}
where $l,m\in\bbZ$, which from \eqref{eqn:conds_f_vari_A4_2} satisfy
\begin{equation}\label{eqn:cond_f_fun_A4}
 f_{l,m}^{(5)}f_{l,m}^{(1)}=\dfrac{a_0a_1(q^{-m}a_3+q^la_0f_{l,m}^{(3)})}{q^{-m}a_2{a_3}^2}.
\end{equation}
The $f$-functions satisfy the $q$-difference equation for $l$-direction:
\begin{equation}\label{eqn:A4_qP_T0}
 \begin{cases}
 f_{l+1,m}^{(3)}f_{l,m}^{(3)}
 =\dfrac{q^{-m}a_{3}(q^{-l}a_{1}+q^{-m}a_{3}a_{4}f_{l,m}^{(1)})(q^{-l}a_{1}+q^{-m}a_{3}f_{l,m}^{(1)})}{{a_0}^2{a_{1}}^2a_{4}(a_0a_{1}+q^{-m}a_{3}f_{l,m}^{(1)})},\\[1em]
 f_{l+1,m}^{(1)}f_{l,m}^{(1)}
 =\dfrac{q^{-2l-2}a_0{a_{1}}^3(a_{2}a_{3}+q^{l+1}a_0f_{l+1,m}^{(3)})(q^{-m}a_{3}+q^{l+1}a_0f_{l+1,m}^{(3)})}{q^{-2m}{a_{3}}^2(q^{-m}a_{3}+a_0a_{1}f_{l+1,m}^{(3)})},
 \end{cases}
\end{equation}
and that for $m$-direction:
\begin{equation}\label{eqn:A4_qP_T2}
 \begin{cases}
 f_{l,m+1}^{(5)}f_{l,m}^{(5)}
 =\dfrac{q^la_{0}(q^{-m}a_{3}+a_{0}a_{1}f_{l,m}^{(3)})(q^{-m}a_{3}+q^la_{0}f_{l,m}^{(3)})}{q^{-l}{a_2}^2{a_{3}}^2a_{1}(a_2a_{3}+q^la_{0}f_{l,m}^{(3)})},\\[1em]
 f_{l,m+1}^{(3)}f_{l,m}^{(3)}
 =\dfrac{q^{-2m-2}a_2{a_{3}}^3(q^la_{4}a_{0}+q^{m+1}a_2f_{l,m+1}^{(5)})(q^la_{0}+q^{m+1}a_2f_{l,m+1}^{(5)})}{q^{2l}{a_{0}}^2(q^la_{0}+a_2a_{3}f_{l,m+1}^{(5)})}.
 \end{cases}
\end{equation}
Note that Equations \eqref{eqn:A4_qP_T0} and \eqref{eqn:A4_qP_T2} follow from
\begin{align}
 &\begin{cases}
  T_1(f_1^{(3)})f_1^{(3)}
 =\dfrac{a_{3}(a_{1}+a_{3}a_{4}f_1^{(1)})(a_{1}+a_{3}f_1^{(1)})}
 {{a_0}^2{a_{1}}^2a_{4}(a_0a_{1}+a_{3}f_1^{(1)})},\\[1em]
 {T_1}^{-1}(f_1^{(1)})f_1^{(1)}
 =\dfrac{a_0{a_{1}}^3(a_{2}a_{3}+a_0f_1^{(3)})(a_{3}+a_0f_1^{(3)})}{{a_{3}}^2(a_{3}+a_0a_{1}f_1^{(3)})},
 \end{cases}\\
&\begin{cases}
 T_2(f_1^{(5)})f_1^{(5)}
 =\dfrac{a_{0}(a_{3}+a_{0}a_{1}f_1^{(3)})(a_{3}+a_{0}f_1^{(3)})}
 {{a_2}^2{a_{3}}^2a_{1}(a_2a_{3}+a_{0}f_1^{(3)})},\\[1em]
 {T_2}^{-1}(f_1^{(3)})f_1^{(3)}
 =\dfrac{a_2{a_{3}}^3(a_{4}a_{0}+a_2f_1^{(5)})(a_{0}+a_2f_1^{(5)})}{{a_{0}}^2(a_{0}+a_2a_{3}f_1^{(5)})},
\end{cases}
\end{align}
respectively.
\begin{remark}
Each of equations \eqref{eqn:A4_qP_T0} and \eqref{eqn:A4_qP_T2} is equivalent to the \qP{V} \eqref{eqn:qP_appendix_A4}.
The correspondence between Equations \eqref{eqn:A4_qP_T0} and \eqref{eqn:qP_appendix_A4} is given by
\begin{equation}
\begin{split}
 &F=f_{l,m}^{(1)},\quad
 G=f_{l,m}^{(3)},\quad
 \overline{F}=f_{l+1,m}^{(1)},\quad
 \overline{G}=f_{l+1,m}^{(3)},\\
 &t=q^{l-m} {a_1}^{-1}{a_2}^{-1},\quad
 c_1=q^{-1}a_0a_1,\quad
 c_2={a_2}^{-1}{a_3}^{-1},\quad
 c_3=q^ma_0a_1{a_3}^{-1},
\end{split}
\end{equation}
while that between Equations \eqref{eqn:A4_qP_T2} and \eqref{eqn:qP_appendix_A4} is given by
\begin{equation}
\begin{split}
 &F=f_{l,m}^{(3)},\quad
 G=f_{l,m}^{(5)},\quad
 \overline{F}=f_{l,m+1}^{(3)},\quad
 \overline{G}=f_{l,m+1}^{(5)},\\
 &t=q^{m} {a_3}^{-1}{a_4}^{-1},\quad
 c_1=q^{-1}a_2a_3,\quad
 c_2=q^{-l}{a_4}^{-1}{a_0}^{-1},\quad
 c_3=q^{-l}a_2a_3{a_0}^{-1}.
\end{split} 
\end{equation}
\end{remark}

Letting
\begin{equation}\label{eqn:A4_u_qPA4}
 u=\dfrac{(1-a_1a_2)a_2{a_3}^{3/2}f_1^{(5)}}{q^{1/4}{a_0}^{1/2}(a_3+a_0a_1f_1^{(3)})},
\end{equation}
we can verify that the following relation holds:
\begin{equation}
 T_1T_2(u)-u=\dfrac{qa_2-{a_1}^{-1}}{T_2(u)}-\dfrac{a_2-q{a_1}^{-1}}{T_1(u)}.
\end{equation}
Therefore, applying ${T_1}^l{T_2}^m$ to the equation above
we obtain
\begin{equation}
 U_{l+1,m+1}-U_{l,m}
 =\dfrac{q^{m+1}a_2-q^l{a_1}^{-1}}{U_{l,m+1}}-\dfrac{q^ma_2-q^{l+1}{a_1}^{-1}}{U_{l+1,m}},
\end{equation}
where $U_{l,m}={T_1}^{l}{T_2}^{m}(u)$, which is equivalent to the multiplicative dKdV equation \eqref{eqn:mul_dKdV_1} with the following correspondence:
\begin{equation}\label{eqn:corre_dKdV_A4fun}
 u_{l,m}=U_{l,m},\quad
 \al_{l}=q^{l}{a_1}^{-1},\quad
 \be_{m}=q^{m}a_2,\quad
 \ep=q.
\end{equation}
Therefore, setting
\begin{equation}
\begin{split}
 &d_1={a_0}^{1/2}{a_1}^{1/2},\quad
 d_2={a_2}^{1/2}{a_3}^{1/2},\quad
 f_{l,m}={d_2}^2f_{l,m}^{(1)},\quad
 g_{l,m}={d_1}^2{d_2}^{-2}f_{l,m}^{(3)},\\
 &h_{l,m}={d_2}^2f_{l,m}^{(5)},
\end{split}
\end{equation}
and using Equations \eqref{eqn:cond_f_fun_A4}, \eqref{eqn:A4_qP_T0}, \eqref{eqn:A4_qP_T2} and \eqref{eqn:A4_u_qPA4} with the correspondence \eqref{eqn:corre_dKdV_A4fun},
we have completed the proof of Theorem \ref{theorem:A4}.

\subsection{Proof of Theorem \ref{theorem:A3}}\label{subsection:A3_proof}
In this subsection using the transformation group $\widetilde{W}(D_5^{(1)})$ we give a proof of Theorem \ref{theorem:A3}.

Let $a_i$ $(i=0,\dots,5)$ and $q$ be complex parameters satisfying
\begin{equation}
 a_0a_1{a_2}^2{a_3}^2a_4a_5=q,
\end{equation}
and $f_0$, $f_1$ be complex variables.
The transformation group $\widetilde{W}(D_5^{(1)})=\langle s_0,\dots,s_5,\sigma_1,\sigma_2\rangle$ is defined by the actions on the parameters:
\begin{align*}
 &s_i(a_j)=a_j{a_i}^{-C_{ij}},\\
 &\sigma_1:(a_0,a_1,a_2,a_3,a_4,q)\mapsto({a_5}^{-1},{a_4}^{-1},{a_3}^{-1},{a_2}^{-1},{a_1}^{-1},{a_0}^{-1},q^{-1}),\\
 &\sigma_2:(a_0,a_1,a_2,a_3,a_4,q)\mapsto({a_1}^{-1},{a_0}^{-1},{a_2}^{-1},{a_3}^{-1},{a_4}^{-1},{a_5}^{-1},q^{-1}),
\end{align*}
where $i,j\in\bbZ/6\bbZ$ and $(C_{ij})_{i,j=0}^5$ is the Cartan matrix of type $D_5^{(1)}$:
\begin{equation*}
 (C_{ij})_{i,j=0}^5
 =\left(\begin{array}{cccccc}
 2&0&-1&0&0&0\\0&2&-1&0&0&0\\-1&-1&2&-1&0&0\\0&0&-1&2&-1&-1\\0&0&0&-1&2&0\\0&0&0&-1&0&2
 \end{array}\right)
\end{equation*}
and those on the variables: 
\begin{align*}
 &s_2(f_1)=\dfrac{f_1(a_0f_0+a_1{a_2}^2)}{a_0{a_2}^2f_0+a_1},\quad
 s_3(f_0)=\dfrac{f_0({a_3}^2a_4f_1+a_5)}{a_4f_1+{a_3}^2a_5},\\
 &\sigma_1(f_0)=f_1,\quad
 \sigma_1(f_1)=f_0,\quad
 \sigma_2(f_1)=\dfrac{1}{f_1}.
\end{align*}
See Remark \ref{remark:action_identity} for the convention on how to write these actions.
The transformation group $\widetilde{W}(D_5^{(1)})$ satisfies the fundamental relations of the extended affine Weyl group of type $D_5^{(1)}$\cite{SakaiH2001:MR1882403,TM2006:MR2202304,TakenawaT2003:MR1996297}:
\begin{subequations}
\begin{align}
 &{s_i}^2=1,\quad
 (s_is_j)^2=1\quad (\text{if}~C_{ij}=0),\quad
 (s_is_j)^3=1\quad (\text{if}~C_{ij}=-1),\\
 &{\sigma_1}^2={\sigma_2}^2=(\sigma_1\sigma_2)^4=1,\quad
 \sigma_1s_{\{0,1,2,3,4,5\}}=s_{\{5,4,3,2,1,0\}}\sigma_1,\\
 &\sigma_2s_{\{0,1,2,3,4,5\}}=s_{\{1,0,2,3,4,5\}}\sigma_2,\quad
\end{align}
\end{subequations}
where $i,j\in\bbZ/6\bbZ$.

Define the translations $T_1$ and $T_2$ by
\begin{equation}
 T_1=(\sigma_2\sigma_1 s_0s_2s_3s_4)^2,\quad
 T_2=(\sigma_2\sigma_1 s_5s_3s_2s_0)^2,
\end{equation}
whose actions on the parameters are given by
\begin{subequations}
\begin{align}
 &T_1:(a_0,a_1,a_2,a_3,a_4,a_5)\mapsto(a_0,a_1,a_2,a_3,q^{-1}a_4,qa_5),\\
 &T_2:(a_0,a_1,a_2,a_3,a_4,a_5)\mapsto(q^{-1}a_0,qa_1,a_2,a_3,a_4,a_5).
\end{align}
\end{subequations}
Note that the translations $T_1$ and $T_2$ commute with each other and the parameter $q$ is invariant under the action of each translation.

Let us introduce the additional variables $g_0$ and $g_1$ by
\begin{equation}\label{eqn:g_vari_A3}
 g_0=\dfrac{({a_0}^3a_1{a_2}^2+f_0)f_1}{a_0(a_1+a_0{a_2}^2f_0)},\quad
 g_1=\dfrac{{a_3}^2{a_4}^3a_5+f_1}{a_4(a_5+{a_3}^2a_4f_1)f_0}.
\end{equation}
Moreover, we define 
\begin{equation}
 f_i^{l,m}={T_1}^l{T_2}^m(f_i),\quad
 g_i^{l,m}={T_1}^l{T_2}^m(g_i),\quad
 (i=0,1)
\end{equation}
where $l,m\in\bbZ$, which from \eqref{eqn:g_vari_A3} satisfy
\begin{equation}\label{eqn:cond_fg_fun_A3}
 g_0^{l,m}=\dfrac{(q^{-2m}{a_0}^3a_1{a_2}^2+f_0^{l,m})f_1^{l,m}}{q^{-m}a_0(q^ma_1+q^{-m}a_0{a_2}^2f_0^{l,m})},\quad
 g_1^{l,m}=\dfrac{q^{-2l}{a_3}^2{a_4}^3a_5+f_1^{l,m}}{q^{-l}a_4(q^la_5+q^{-l}{a_3}^2a_4f_1^{l,m})f_0^{l,m}}.
\end{equation}
These functions satisfy the $q$-difference equation for $l$-direction:
\begin{equation}\label{eqn:A3_qP_T1}
 \begin{cases}
 g_1^{l+1,m}g_1^{l,m}
 =\dfrac{{a_4}^2(1+q^{2l}{a_3}^2a_4{a_5}^3f_1^{l,m})(1+q^{2l}{a_3}^{-2}{a_4}^{-3}{a_5}^{-1}f_1^{l,m})}
 {(q^{2l}a_5+{a_3}^{-2}a_4f_1^{l,m})(q^{2l}a_5+{a_3}^2a_4f_1^{l,m})},\\[1em]
 f_1^{l+1,m}f_1^{l,m}
 =\dfrac{{a_4}^2(1+q^{2l+2m+1}{a_1}^{2}{a_4}^{-1}a_5g_1^{l+1,m})(1+q^{2l-2m+1}{a_1}^{-2}{a_4}^{-1}a_5g_1^{l+1,m})}
 {(q^{2l+1}a_5+q^{2m}{a_0}^{-2}a_4g_1^{l+1,m})(q^{2l+1}a_5+q^{-2m}{a_0}^2a_4g_1^{l+1,m})},
 \end{cases}
\end{equation}
and that for $m$-direction:
\begin{equation}\label{eqn:A3_qP_T2}
 \begin{cases}
  g_0^{l,m+1}g_0^{l,m}
  =\dfrac{{a_0}^2(1+q^{2m}a_0{a_1}^3{a_2}^2f_0^{l,m})(1+q^{2m}{a_0}^{-3}{a_1}^{-1}{a_2}^{-2}f_0^{l,m})}
  {(q^{2m}a_1+a_0{a_2}^{-2}f_0^{l,m})(q^{2m}a_1+a_0{a_2}^2f_0^{l,m})},\\[1em]
 f_0^{l,m+1}f_0^{l,m}
 =\dfrac{{a_0}^2(1+q^{2m-2l+1}{a_0}^{-1}a_1{a_4}^2g_0^{l,m+1})(1+q^{2m+2l+1}{a_0}^{-1}a_1{a_4}^{-2}g_0^{l,m+1})}
 {(q^{2m+1}a_1+q^{-2l}a_0{a_5}^{-2}g_0^{l,m+1})(q^{2m+1}a_1+q^{2l}a_0{a_5}^2g_0^{l,m+1})}.
 \end{cases}
\end{equation}
Note that Equations \eqref{eqn:A3_qP_T1} and \eqref{eqn:A3_qP_T2} follow from
\begin{align}
 &\begin{cases}
 T_1(g_1)g_1=\dfrac{(1+{a_3}^2a_4{a_5}^3f_1)({a_3}^2{a_4}^3a_5+f_1)}{a_4a_5({a_3}^2a_5+a_4f_1)(a_5+{a_3}^2a_4f_1)},\\[1em]
 {T_1}^{-1}(f_1)f_1=\dfrac{a_0(a_0{a_2}^2{a_3}^2{a_4}^2+a_1g_1)(a_0{a_1}^3{a_2}^2{a_3}^2{a_4}^2+g_1)}{a_1(a_0+a_1{a_2}^2{a_3}^2{a_4}^2g_1)(1+{a_0}^3a_1{a_2}^2{a_3}^2{a_4}^2g_1)},
 \end{cases}\\[0.5em]
&\begin{cases}
 T_2(g_0)g_0=\dfrac{(1+a_0{a_1}^3{a_2}^2f_0)({a_0}^3a_1{a_2}^2+f_0)}{a_0a_1(a_1{a_2}^2+a_0f_0)(a_1+a_0{a_2}^2f_0)},\\[1em]
 {T_2}^{-1}(f_0)f_0=\dfrac{a_5({a_0}^2{a_2}^2{a_3}^2a_5+a_4g_0)({a_0}^2{a_2}^2{a_3}^2{a_4}^3a_5+g_0)}{a_4(a_5+{a_0}^2{a_2}^2{a_3}^2a_4g_0)(1+{a_0}^2{a_2}^2{a_3}^2a_4{a_5}^3g_0)},
\end{cases}
\end{align}
respectively.
\begin{remark}
Each of equations \eqref{eqn:A3_qP_T1} and \eqref{eqn:A3_qP_T2} is equivalent to the \qP{VI} \eqref{eqn:qP_appendix_A3}.
The correspondence between Equations \eqref{eqn:A3_qP_T1} and \eqref{eqn:qP_appendix_A3} is given by
\begin{equation}
\begin{split}
 &F=\dfrac{a_4}{q^{2l}a_5}f_1^{l,m},\quad
 G=\dfrac{a_4}{q^{2l-1}a_5}g_1^{l,m},\quad
 \overline{F}=\dfrac{a_4}{q^{2l+2}a_5}f_1^{l+1,m},\quad
 \overline{G}=\dfrac{a_4}{q^{2l+1}a_5}g_1^{l+1,m},\\
 &t=\dfrac{q^l{a_5}^{1/2}}{{a_4}^{1/2}},\quad
 c_1=(a_3a_4a_5)^{2},\quad
 c_2={a_3}^{2},\quad
 c_3=q^{2m}{a_1}^{2},\quad
 c_4=q^{-2m}{a_0}^{2},
\end{split}
\end{equation}
while that between Equations \eqref{eqn:A3_qP_T2} and \eqref{eqn:qP_appendix_A3} is given by
\begin{equation}
\begin{split}
 &F=\dfrac{a_0}{q^{2m}a_1}f_0^{l,m},\quad
 G=\dfrac{a_0}{q^{2m-1}a_1}g_0^{l,m},\quad
 \overline{F}=\dfrac{a_0}{q^{2m+2}a_1}f_0^{l,m+1},\quad
 \overline{G}=\dfrac{a_0}{q^{2m+1}a_1}g_0^{l,m+1},\\
 &t=\dfrac{q^m{a_1}^{1/2}}{{a_0}^{1/2}},\quad
 c_1=(a_0a_1a_2)^{2},\quad
 c_2={a_2}^{2},\quad
 c_3=q^{-2l}{a_4}^{2},\quad
 c_4=q^{2l}{a_5}^{2}.
\end{split}
\end{equation}
\end{remark}

Letting
\begin{equation}\label{eqn:A3_u_qPA3}
 u=\cfrac{a_5({a_1}^4{a_2}^4{a_3}^4{a_4}^4-1)\left(a_1{a_2}^2{a_3}^2+\dfrac{a_4(a_1{a_2}^2+a_0f_0)f_1}{a_5(1+a_0{a_1}^3{a_2}^2f_0)}\right)}{a_1{a_2}^2{a_3}^2a_4f_0\left(1+\dfrac{{a_1}^3{a_2}^2{a_3}^2a_4(a_1{a_2}^2+a_0f_0)f_1}{a_5(1+a_0{a_1}^3{a_2}^2f_0)}\right)},
\end{equation}
we can verify that the following relation holds:
\begin{equation}
\begin{split}
 T_1T_2(u)-u
 =&\dfrac{1-q^4{a_1}^4{a_2}^4{a_3}^4({a_4}^4+{a_5}^4-q^4{a_1}^4{a_2}^4{a_3}^4{a_4}^4{a_5}^4)}{q^4{a_2}^4{a_3}^4{a_4}^4T_2(u)}\\
 &+\dfrac{{a_4}^4(1-{a_0}^4{a_2}^4{a_3}^4{a_5}^4)(q^{-4}-{a_1}^4{a_2}^4{a_3}^4{a_5}^4)}{T_1(u)}.
\end{split}
\end{equation}
Therefore, applying ${T_1}^l{T_2}^m$ to the equation above
we obtain
\begin{equation}\label{eqn:A3_U_dKdV}
 U_{l+1,m+1}-U_{l,m}
 =\dfrac{B_{m+1}-A_l}{U_{l,m+1}}-\dfrac{B_{m}-A_{l+1}}{U_{l+1,m}},
\end{equation}
where $U_{l,m}={T_1}^{l}{T_2}^{m}(u)$ and
\begin{align}
 &A_l=q^{-4l}{a_5}^{-4}(1-q^{4l}{a_5}^4)({a_4}^4{a_5}^4-q^{4l}{a_5}^4),\\
 &B_m=q^{-4m}{a_1}^{-4}{a_2}^{-4}{a_3}^{-4}(1-q^{4m}{a_1}^4{a_2}^4{a_3}^4)(1-q^{4m}{a_1}^4{a_2}^4{a_3}^4{a_4}^4{a_5}^4).
\end{align}
Equation \eqref{eqn:A3_U_dKdV} is equivalent to the multiplicative dKdV equation \eqref{eqn:mul_dKdV_2} and the correspondence is given by the following:
\begin{equation}\label{eqn:corre_dKdV_A3fun}
 u_{l,m}=U_{l,m},\quad
 \al_{l}=q^{4l}{a_5}^4,\quad
 \be_{m}=q^{4m}({a_1}{a_2}{a_3})^4,\quad
 \ga=({a_4}{a_5})^4,\quad
 \ep=q^4.
\end{equation}
Therefore, setting
\begin{equation}
\begin{split}
 &d_1={a_2}^2,\quad
 d_2=({a_2}{a_3})^4,\quad
 f_{l,m}=\dfrac{\ga^{1/4}f_1^{l,m}}{{\al_l}^{1/2}},\quad
 g_{l,m}=\dfrac{\ep^{1/4}\ga^{1/4}g_1^{l,m}}{{\al_l}^{1/2}},\\
 &x_{l,m}=\dfrac{\ep^{1/4}f_0^{l,m}}{{\be_m}^{1/2}\ga^{1/4}},\quad
 y_{l,m}=\dfrac{\ep^{1/2}g_0^{l,m}}{{\be_m}^{1/2}\ga^{1/4}},
\end{split}
\end{equation}
and using Equations \eqref{eqn:cond_fg_fun_A3}, \eqref{eqn:A3_qP_T1}, \eqref{eqn:A3_qP_T2} and \eqref{eqn:A3_u_qPA3} with the correspondence \eqref{eqn:corre_dKdV_A3fun},
we have completed the proof of Theorem \ref{theorem:A3}.
\section{Concluding remarks}\label{ConcludingRemarks}
In this paper, we have constructed the special solutions to the multiplicative dKdV equations \eqref{eqn:mul_dKdV_1} and \eqref{eqn:mul_dKdV_2}.
The distinctive feature of these solutions is that along each direction for $l\in\bbZ$ and $m\in\bbZ$ they are represented by discrete Painlev\'e transcendents (see Theorems \ref{theorem:A6}--\ref{theorem:A3}).
Although not explicitly mentioned, special solutions with similar features for another 2-dimensional lattice equation can be found in \cite{NakazonoN2018:MR3760161}.
In \cite{NakazonoN2018:MR3760161}, each of such solutions is constructed by imposing a periodic condition on a system of 2-dimensional lattice equations. 
It is a matter of future work to consider what kind of constraint is imposed on the special solutions obtained in this paper. 

All the $q$-Painlev\'e equations treated in this paper arise from the translations of the corresponding extended affine Weyl groups. 
However, $q$-Painlev\'e equations and the lmKdV equation arise not only from translations but also arise from non-translation elements\cite{KNT2011:MR2773334,JNS2016:MR3584386,JNS2015:MR3403054}.
This implies that we can also construct discrete Painlev\'e transcendent solutions to the multiplicative type dKdV equations using non-translations. 
The results in this direction will be reported in forthcoming publications.
\subsection*{Acknowledgment}
This research was supported by a JSPS KAKENHI Grant Number JP19K14559.
\appendix
\section{$q$-Painlev\'e equations of types $A_3^{(1)}$, $A_4^{(1)}$, $A_5^{(1)}$ and $A_6^{(1)}$}\label{section:typical_examples_dPs}
We here list some typical examples of $q$-Painlev\'e equations of types $A_J^{(1)}$ ($J=3,4,5,6$).
Note that in the following
$t\in\bbC^\ast$ plays the role of an independent variable,
$F(t),G(t)\in\bbC$ play the roles of dependent variables
and $c_i,q\in\bbC^\ast$ play the roles of parameters.
Moreover, we adopt the following shorthand notations for the dependent variables:
\begin{equation}
 F=F(t),\quad
 G=G(t),\quad
 \overline{F}=F(q t),\quad
 \overline{G}=G(q t),\quad
 \underline{F}=F(q^{-1} t).
\end{equation}
\begin{itemize}
\item[\underline{$A_6^{(1)}$-type}]
\begin{align}
 \qP{II}:~&~\left(\overline{F}\,F-\dfrac{1}{t}\right)\left(\underline{F}\,F-\dfrac{q}{t}\right)=\dfrac{c_1 F}{t(F+1)}
 \label{eqn:qp2_appendix_A6}\\[0.5em]
 \qPD:~&~
 \overline{G}G=\dfrac{1+t^{-1}F}{F(1+{c_1}^{-1}F)},\quad
 \overline{F}F=\dfrac{c_1(1+\overline{G})}{\overline{G}^2}
 \label{eqn:qp_appendix_A6_2}
\end{align}
\item[\underline{$A_5^{(1)}$-type}]
\begin{equation}\label{eqn:qp3_appendix_A5}
 \qP{III}:~\overline{G}G=\dfrac{c_1(1+t F)}{F(t+F)},\quad
 \overline{F}F=\dfrac{c_1(1+c_2 t \overline{G})}{\overline{G}(c_2 t+\overline{G})}
\end{equation}
\item[\underline{$A_4^{(1)}$-type}]
\begin{equation}\label{eqn:qP_appendix_A4}
 \qP{V}:~
 \begin{cases}
 \,\overline{G}G=\dfrac{(c_1+t F)(c_2+t F)}{{c_3}^2t^2({c_3}+F)},\\[1em]
 \,\overline{F}F=\dfrac{{c_3}^2({c_3}^{-1}+q t \overline{G})(c_1c_2{c_3}^{-2}+t\overline{G})}{qt^2({c_3}^{-1}+\overline{G})}
 \end{cases}
\end{equation}
\item[\underline{$A_3^{(1)}$-type}]
\begin{equation}\label{eqn:qP_appendix_A3}
 \qP{VI}:~
 \begin{cases}
 \,\overline{G}G=\dfrac{(F+c_1t^{-4})(F+{c_1}^{-1}t^{-4})}{(F+c_2)(F+{c_2}^{-1})},\\[1em]
 \,\overline{F}F=\dfrac{(\overline{G}+q^{-2}c_3t^{-4})(\overline{G}+q^{-2}{c_3}^{-1}t^{-4})}{(\overline{G}+c_4)(\overline{G}+{c_4}^{-1})}
 \end{cases}
\end{equation}
\end{itemize}

\begin{remark}
Equations \eqref{eqn:qp2_appendix_A6}--\eqref{eqn:qP_appendix_A3} are known as  a $q$-discrete analogue of the Painlev\'e II equation\cite{RG1996:MR1399286}, that of the Painlev\'e III equation of type $D_7^{(1)}$\cite{RGTT2000}, that of the Painlev\'e III equation \cite{CNP1991:MR1111648}, that of the Painlev\'e V equation \cite{SakaiH2001:MR1882403} and that of the Painlev\'e VI equation \cite{JS1996:MR1403067}, respectively.
\end{remark}

The following is a list of correspondences between the $q$-Painlev\'e equations \eqref{eqn:qp_appendix_A6_2}--\eqref{eqn:qP_appendix_A3} and those in Theorems \ref{theorem:A6}--\ref{theorem:A3}.
\begin{description}
\item[Equations \eqref{eqn:qp_appendix_A6_2} and \eqref{eqn:theo_A6_l-direction}]
\begin{equation}
\begin{split}
 &F=h_{l,m},\quad
 G=g_{l-1,m},\quad
 \overline{F}=h_{l+1,m},\quad
 \overline{G}=g_{l,m},\\
 &t=\al_{l+2},\quad
 q=\ep,\quad
 c_1=\be_{m+2}.
\end{split}
\end{equation}
\item[Equations \eqref{eqn:qp_appendix_A6_2} and \eqref{eqn:theo_A6_m-direction}]
\begin{equation}
\begin{split}
 &F=h_{l,m},\quad
 G=f_{l,m},\quad
 \overline{F}=h_{l,m+1},\quad
 \overline{G}=f_{l,m+1},\\
 &t=\be_{m+2},\quad
 q=\ep,\quad
 c_1=\al_{l+2}.
\end{split}
\end{equation}
\item[Equations \eqref{eqn:qp3_appendix_A5} and \eqref{eqn:theo_A5_l-direction}]
\begin{equation}
\begin{split}
 &F=f_{l,m},\quad
 G=g_{l,m},\quad
 \overline{F}=f_{l+1,m},\quad
 \overline{G}=g_{l+1,m},\\
 &t={\al_{l}}^{-1/2},\quad
 q=\ep^{-1/2},\quad
 c_1=d^2,\quad
 c_2={\be_{m-1}}^{1/2}.
\end{split}
\end{equation}
\item[Equations \eqref{eqn:qp3_appendix_A5} and \eqref{eqn:theo_A5_m-direction}]
\begin{equation}
\begin{split}
 &F=g_{l,m},\quad
 G=h_{l,m},\quad
 \overline{F}=g_{l,m+1},\quad
 \overline{G}=h_{l,m+1},\\
 &t={\al_{l}}^{1/2}{\be_{m}}^{-1/2},\quad
 q=\ep^{-1/2},\quad
 c_1=d^2,\quad
 c_2={\al_{l}}^{-1/2}.
\end{split}
\end{equation}
\item[Equations \eqref{eqn:qP_appendix_A4} and \eqref{eqn:theo_A4_l-direction}]
\begin{equation}
\begin{split}
 &F=\dfrac{f_{l,m}}{{d_2}^2},\quad
 G=\dfrac{{d_2}^2g_{l,m}}{{d_1}^2},\quad
 \overline{F}=\dfrac{f_{l+1,m}}{{d_2}^2},\quad
 \overline{G}=\dfrac{{d_2}^2g_{l+1,m}}{{d_1}^2},\\
 &t=\dfrac{\al_l}{\be_m},\quad
 q=\ep,\quad
 c_1=\dfrac{{d_1}^2}{\ep},\quad
 c_2=\dfrac{1}{{d_2}^2},\quad
 c_3=\dfrac{{d_1}^2\be_m}{{d_2}^2}.
\end{split}
\end{equation}
\item[Equations \eqref{eqn:qP_appendix_A4} and \eqref{eqn:theo_A4_m-direction}]
\begin{equation}
\begin{split}
 &F=\dfrac{{d_2}^2g_{l,m}}{{d_1}^2},\quad
 G=\dfrac{h_{l,m}}{{d_2}^2},\quad
 \overline{F}=\dfrac{{d_2}^2g_{l,m+1}}{{d_1}^2},\quad
 \overline{G}=\dfrac{h_{l,m+1}}{{d_2}^2},\\
 &t={d_1}^2\be_{m-1},\quad
 q=\ep,\quad
 c_1=\dfrac{{d_2}^2}{\ep},\quad
 c_2=\dfrac{{d_2}^2}{\al_{l+1}},\quad
 c_3=\dfrac{{d_2}^2}{{d_1}^2\al_l}.
\end{split}
\end{equation}
\item[Equations \eqref{eqn:qP_appendix_A3} and \eqref{eqn:theo_A3_l-direction}]
\begin{equation}
\begin{split}
 &F=f_{l,m},\quad
 G=g_{l,m},\quad
 \overline{F}=f_{l+1,m},\quad
 \overline{G}=g_{l+1,m},\quad
 t=\dfrac{{\al_l}^{1/4}}{\ga^{1/8}},\quad
 q=\ep^{1/4},\\
 &c_1=\dfrac{{d_2}^{1/2}\ga^{1/2}}{d_1},\quad
 c_2=\dfrac{{d_2}^{1/2}}{d_1},\quad
 c_3=\dfrac{{\be_m}^{1/2}}{{d_2}^{1/2}},\quad
 c_4=\dfrac{\ep^{1/2}}{{d_2}^{1/2}{\be_m}^{1/2}\ga^{1/2}}.
\end{split}
\end{equation}
\item[Equations \eqref{eqn:qP_appendix_A3} and \eqref{eqn:theo_A3_m-direction}]
\begin{equation}
\begin{split}
 &F=x_{l,m}\quad
 G=y_{l,m},\quad
 \overline{F}=x_{l,m+1},\quad
 \overline{G}=y_{l,m+1},\quad
 t=\dfrac{{\be_m}^{1/4}\ga^{1/8}}{\ep^{1/8}},\quad
 q=\ep^{1/4},\\
 &c_1=\dfrac{\ep^{1/2}d_1}{{d_2}\ga^{1/2}},\quad
 c_2=d_1,\quad
 c_3=\dfrac{\ga^{1/2}}{{\al_l}^{1/2}},\quad
 c_4={\al_l}^{1/2}.
\end{split}
\end{equation}
\end{description}

\def\cprime{$'$} \def\cprime{$'$}

\end{document}